\documentclass[12pt]{article}

\usepackage{amsmath}
\usepackage{amssymb}
\usepackage{amsfonts}
\usepackage{mathrsfs}
\usepackage{graphicx}
\usepackage{enumerate}

\usepackage[margin=1in]{geometry}
\usepackage[width=0.9\textwidth]{caption}

\begin{document}

\title{Complexity of the AdS Soliton}
\author{Alan P. Reynolds\footnote{a.p.reynolds@durham.ac.uk}\,\,  and Simon F. Ross\footnote{s.f.ross@durham.ac.uk} \\  \bigskip \\ Centre for Particle Theory, Department of Mathematical Sciences \\ Durham University\\ South Road, Durham DH1 3LE}

\maketitle
 
\begin{abstract}
We consider the holographic complexity conjectures in the context of the AdS soliton, which is the holographic dual of the ground state of a field theory on a torus with antiperiodic boundary conditions for fermions on one cycle. The complexity is a non-trivial function of the size of the circle with antiperiodic boundary conditions, which sets an IR scale in the dual geometry. We find qualitative differences between the calculations of complexity from spatial volume and action (CV and CA). In the CV calculation, the complexity for antiperiodic boundary conditions is smaller than for periodic, and decreases monotonically with increasing IR scale. In the CA calculation,  the complexity for antiperiodic boundary conditions is larger than for periodic, and initially increases with increasing IR scale, eventually decreasing to zero as the IR scale becomes of order the UV cutoff. We compare these results to a simple calculation for free fermions on a lattice, where we find the complexity for antiperiodic boundary conditions is larger than for periodic.
\end{abstract}
 
\clearpage 

\section{Introduction}
 
 In principle, holography provides a well-defined non-perturbative formulation of quantum gravity. But to really use it to address questions about the nature of spacetime, we need to understand the emergence of the bulk spacetime from the dual field theory description. Since the conjecture of Ryu and Takayanagi \cite{Ryu:2006bv}, there has been growing evidence that entanglement plays an important role, and a variety of tools from quantum information have been applied to understand how spacetime emerges from the field theory. In \cite{Susskind:2014rva}, Susskind conjectured a new relation between the bulk geometry and the dual boundary state, proposing that the time-dependent geometry of the region behind the horizon of an AdS black hole could be related to the complexity of the dual boundary state.\footnote{The quantum computational complexity is a measure of the minimum number of elementary gates needed in a quantum circuit which constructs a given state  starting from a specified simple reference state (see e.g. \cite{Osborne:c}).}
This proposal was refined  in \cite{Stanford:2014jda} into the conjecture that the computational complexity of the boundary state at a given time (on some spacelike slice of the boundary) could be identified with the volume of a maximal volume spacelike slice in the bulk, ending on the given boundary slice. This will be referred to as the CV conjecture. This was further developed in \cite{Susskind:2014jwa,Susskind:2014moa}. 

More recently, it was conjectured that the complexity is related instead to the action of a Wheeler-de Witt patch in the bulk bounded by the given spacelike surface \cite{Brown:2015bva,Brown:2015lvg}. This is referred to as the CA conjecture. An appropriate prescription for calculating the action for a region of spacetime bounded by null surfaces was obtained in \cite{Lehner:2016vdi}. Further related work is \cite{Barbon:2015ria,Brown:2016wib,Couch:2016exn,Yang:2016awy,Chapman:2016hwi}.
 
No derivation of these conjectures, relating them back to the basic AdS/CFT dictionary, has yet been given. They are supported by the relation between the results of the bulk calculation and general expectations for the behaviour of the complexity in a generic quantum system. This evidence comes so far from the study of black hole spacetimes. Both the CV and CA conjectures produce results for the complexity that grow linearly in time at late times, coming from the contributions from the region behind the black hole horizon. This linear growth is supposed to be generic for interacting quantum systems in states of non-maximal complexity \cite{Brown:2015lvg,Brown:2017jil}. Furthermore, the time derivative is simply proportional to the mass of the black hole, which can be interpreted as the energy of the state in the dual theory. This saturates a proposed bound on the growth of the complexity \cite{lloyd,Brown:2015lvg}. (Note however that recent studies of the finite-time growth rate find some violations of this bound \cite{Carmi:2017jqz,Kim:2017qrq}; see \cite{Cottrell:2017ayj} for a discussion of the validity of the bound in a black hole context.) From the bulk point of view, it is highly non-trivial that one obtains simply the mass. This relation to the mass persists in studies of the effects of higher-curvature corrections \cite{Tao:2017fsy,Pan:2016ecg,Alishahiha:2017hwg,Wang:2017uiw,Guo:2017rul}, although recent work on the inclusion of flavour branes finds that the bound is satisfied but no longer saturated \cite{Abad:2017cgl}.
 
 It is interesting to study the conjectures in other spacetimes. In recent work \cite{Reynolds:2017lwq}, we investigated the extension of these calculations to solutions which are asymptotically AdS in a de Sitter slicing, dual to field theories in de Sitter space. For a particular set of $\text{de Sitter} \times S^1$ boundary conditions, there are bubble solutions where the geometry terminates at a finite position in the bulk, where the circle direction closes off smoothly. We found striking differences between the CV and CA calculations for these bubble solutions. 
 
 A simpler context in which similar differences can be seen is the field theory on a torus with antiperiodic boundary conditions for fermions on (at least) one cycle. The ground state for such boundary conditions is dual to the AdS soliton \cite{Horowitz:1998ha}, where the cycle with antiperiodic boundary conditions closes off smoothly in the bulk at a ``bubble", at a radius $r_+$ which is inversely proportional to the size of this cycle. In this paper, we consider the holographic CV and CA calculations of the complexity of this ground state. This is a simple adaptation of our earlier calculations; this geometry arises as a limit of the de Sitter geometries we considered previously. In either case we get a time-independent result, which depends non-trivially on the size of the cycle with antiperiodic boundary conditions, through the dependence on the position of the ``bubble" in the bulk soliton solution. 
 
 We again find striking differences between the CV and CA calculations. The CV calculation for this case is straightforward, and gives a smaller complexity for antiperiodic than for periodic boundary conditions. The complexity for antiperiodic boundary conditions decreases monotonically as the circle radius decreases, bringing the bubble closer to the boundary. For the CA calculation, by contrast, the complexity is larger for antiperiodic than for periodic boundary conditions, and initially increases for decreasing circle radius (while that radius is large compared to the UV cutoff scale). It eventually turns around and decreases, going to zero as the bubble approaches the boundary as one would expect. While the complexity goes to zero as the bubble approaches the boundary in both cases, we find that this involves different powers of the separation in the two cases. 
 
The two proposals for holographic complexity thus give very different answers already in this simple context. We make some first steps towards comparing these holographic calculations to field theory, building on  \cite{Jefferson:2017sdb,Chapman:2017rqy}, where a free boson on a toroidal lattice was considered. We extend these calculations to consider free fermions. We can then consider the change in complexity resulting from changing the fermion boundary conditions from periodic to antiperiodic. We find that in a simple lattice calculation, we get a larger result for the complexity for antiperiodic boundary conditions than for periodic, and the complexity increases as the circle radius decreases. It is worth emphasizing that the calculation we carry out has strong limitations, and an important direction for future work is to refine the field theory calculation and see what effect this has on the behaviour we find. 

In section \ref{review}, we review the holographic complexity conjectures. We discuss the AdS soliton solution in section \ref{solitons}, and carry out the CV calculation. In section \ref{action}, we carry out the CA calculation. In section \ref{lattice}, we consider free fermions on a lattice, and give a calculation of the difference in complexity for the two boundary conditions on the fermions. In section \ref{disc}, we conclude with a brief summary of the results and discussion of future directions.
 
\section{Review of CV and CA}
\label{review}

We first review the two proposals for the holographic calculation of the complexity. In the CV conjecture of \cite{Susskind:2014rva}, the complexity $\mathcal C$ of a pure state $|\Psi \rangle$ of a holographic field theory on some spatial slice $\Sigma$ on the boundary of an asymptotically AdS spacetime is identified with the volume $V$ of the maximal volume codimension one slice $B$ in the bulk having its boundary on $\Sigma$,
\begin{equation}
{\mathcal C_{\textnormal{V}}} \propto \frac{V(B)}{G_{\textnormal{N}} l_{\textnormal{AdS}}}.
\end{equation}
%
This was motivated by the study of the behaviour of Schwarzschild-AdS black hole solutions, where it was found that the volume of the maximal volume slice grows linearly with time, even at late boundary times when other observables have thermalized. The complexity conjecture relates this linear growth to linear growth of the complexity of the dual state, which is expected to continue for exponentially long times in a generic interacting theory, starting from a low-complexity initial state (see \cite{Brown:2017jil} for a recent discussion of this growth of complexity).   The volume of the maximal volume slice has a divergence proportional to the volume of $\Sigma$. 
 
In \cite{Brown:2015bva,Brown:2015lvg}, an alternative CA conjecture was proposed. This identifies the complexity of $|\Psi \rangle$ with the action of the ``Wheeler-de Witt patch", the domain of development of the slice $B$ considered previously. The proposal is that
\begin{equation} \label{ca}
{\mathcal C_{\textnormal{A}}} = \frac{S_{\textnormal{W}}}{\pi \hbar},
\end{equation}
where $S_W$ is the action of the Wheeler-de Witt patch. This proposal has the advantage that the formula is more universal, containing no explicit reference to a bulk length scale. It is also often easier to calculate, as there is no maximisation problem to solve. Finding the Wheeler-de Witt patch for a given boundary slice is easier than finding the maximal volume slice. 

For the black hole solutions, the action of the Wheeler-de Witt patch turns out to also exhibit linear growth in time at late times. In \cite{Brown:2015bva,Brown:2015lvg}, it was argued that the black hole saturates a conjectured  universal upper bound on the rate of growth of the complexity \cite{lloyd}
 \begin{equation} \label{ctd}
 \frac{d \mathcal C}{dt} \leq \frac{2 M}{\pi \hbar}. 
 \end{equation}
From the field theory point of view, the mass $M$ is the energy of the state. The saturation says that black holes represent the situation where the complexity is growing at its maximal rate, analogous to the conjecture that black holes are the fastest scramblers in nature \cite{Hayden:2007cs,Sekino:2008he,Maldacena:2015waa}. Note however that recent studies of the finite-time growth rates in black holes found violations of the bound \cite{Carmi:2017jqz,Kim:2017qrq}. Further work on the CA proposal for charged black holes is found in \cite{Cai:2016xho,Cai:2017sjv,Tao:2017fsy}, while extensions to black holes in more general gravitational theories are found in \cite{Pan:2016ecg,Alishahiha:2017hwg,Wang:2017uiw,Guo:2017rul}. 

This bound is saturated by all Schwarzschild-AdS black holes in the CA conjecture \cite{Brown:2015bva,Brown:2015lvg}. It is also saturated in the CV conjecture for large Schwarzschild-AdS black holes if we take an appropriate normalization of the complexity in the latter case, 
\begin{equation} \label{cvnorm}
\mathcal C_{\textnormal{V}} = \frac{(d-1) V}{2 \pi^2 G_{\textnormal{N}} \ell} = \frac{8 (d-1) V}{\pi \ell}, 
\end{equation}
where in the second equality we adopt units where $16 \pi G_{\textnormal{N}} = 1$, as we will do henceforth. We will adopt this normalization of the complexity in the CV conjecture for definiteness in our later calculations. 

To apply the CA conjecture, we need to define a prescription for the calculation of the action. The Wheeler-de Witt patch has null boundaries, for which the appropriate boundary terms needed for the Einstein-Hilbert action were not yet known. In \cite{Lehner:2016vdi}, motivated by the CA conjecture, a prescription for the action of a region of spacetime containing null boundaries was constructed (see also \cite{Parattu:2015gga,Jubb:2016qzt}). The prescription was obtained by requiring that the variation of the action vanish on-shell when the variation of the metric vanishes on the boundary of the region. 

The prescription of \cite{Lehner:2016vdi} is coordinate-dependent; the value of the action depends on the parametrization adopted on null geodesics along the null boundaries. However, there is a term identified in \cite{Lehner:2016vdi}  which can be added to the action to eliminate this dependence. In previous work \cite{Reynolds:2016rvl}, we showed that including this term has the additional virtue that it removes the leading divergences in the action. 

For the AdS soliton and pure AdS solutions we consider in this paper, the action of the Wheeler-de Witt patch in the prescription of \cite{Lehner:2016vdi} is 
\begin{equation} \label{adsa}
S_{\cal V} = \int_{W} (R- 2 \Lambda) \sqrt{-g} \mathop{dV}  - 2  \int_{F} \kappa \mathop{dS}\mathop{d\lambda}   + 2  \int_{P} \kappa \mathop{dS}\mathop{d\lambda}  -  2   \oint_{\Sigma} a\mathop{dS}, 
\end{equation}
where F (P) are respectively the future (past) boundary of the Wheeler-de Witt patch, $\lambda$ is a parameter on the null generators increasing to the future, so $k^\alpha = \partial x^\alpha /\partial \lambda$ is the tangent to the generators, and $k^\alpha \nabla_\alpha k^\beta = \kappa k^\beta$. $\Sigma$ is their intersection at the AdS boundary, and $a = \ln | k_F \cdot k_P /2|$. 

To this we add the additional contribution
\begin{equation} \label{acorr}
\Delta S = -2 \int_F \Theta \ln | \ell \Theta | dS d\lambda + 2 \int_P \Theta \ln |\ell \Theta | dS d\lambda, 
\end{equation}
where $\Theta = \frac{1}{2} \gamma^{-1} \partial_\lambda \gamma$ is the expansion of the null surfaces, where $\gamma$ is the determinant of the metric on the cross-sections of constant $\lambda$. Our total action is $S = S_{\cal V} + \Delta S$. 

The simplest example of the calculation is to consider vacuum AdS$_{d+1}$ in Poincar\'e coordinates,
\begin{equation} \label{adsp}
ds^2 = \frac{\ell^2}{z^2} ( dz^2 - dt^2 + d\vec{x}^2),
\end{equation}
which is dual to the field theory in flat space. We consider a $d+1$ dimensional AdS space, with a $d$ dimensional boundary. 

For the CV conjecture, the maximal volume slice with boundary at $t=0$ is simply the $t=0$ surface in the bulk, whose volume is 
\begin{equation} \label{vol}
V(B) = \int \mathop{dz}\mathop{d^{d-1}x} \sqrt{h} = \ell^d V_x  \int_\epsilon^\infty \frac{\mathop{dz}}{z^d} = \frac{\ell^d V_x}{(d-1) \epsilon^{d-1}},
\end{equation}
where $V_x$ is the IR divergent coordinate volume in the $\vec x$ directions. Thus, the complexity calculated according to the CV prescription is, with the normalization of \eqref{cvnorm},
\begin{equation} \label{pvol}
\mathcal{C}_{\textnormal{V}} =  \frac{8 \ell^{d-1} V_x}{\pi \epsilon^{d-1}}.
\end{equation}
This is proportional to the volume of the space the field theory lives in, in units of the cutoff.

Turning to the CA conjecture, consider the Wheeler-de Witt patch of this cutoff surface. If we ask for the complexity of the field theory on the $t=0$ surface, cut off at $z = \epsilon$, the Wheeler-de Witt patch lies between $t = z-\epsilon$ and $t = -(z-\epsilon)$.
\begin{figure}
\centering 
\includegraphics[width=0.4\textwidth]{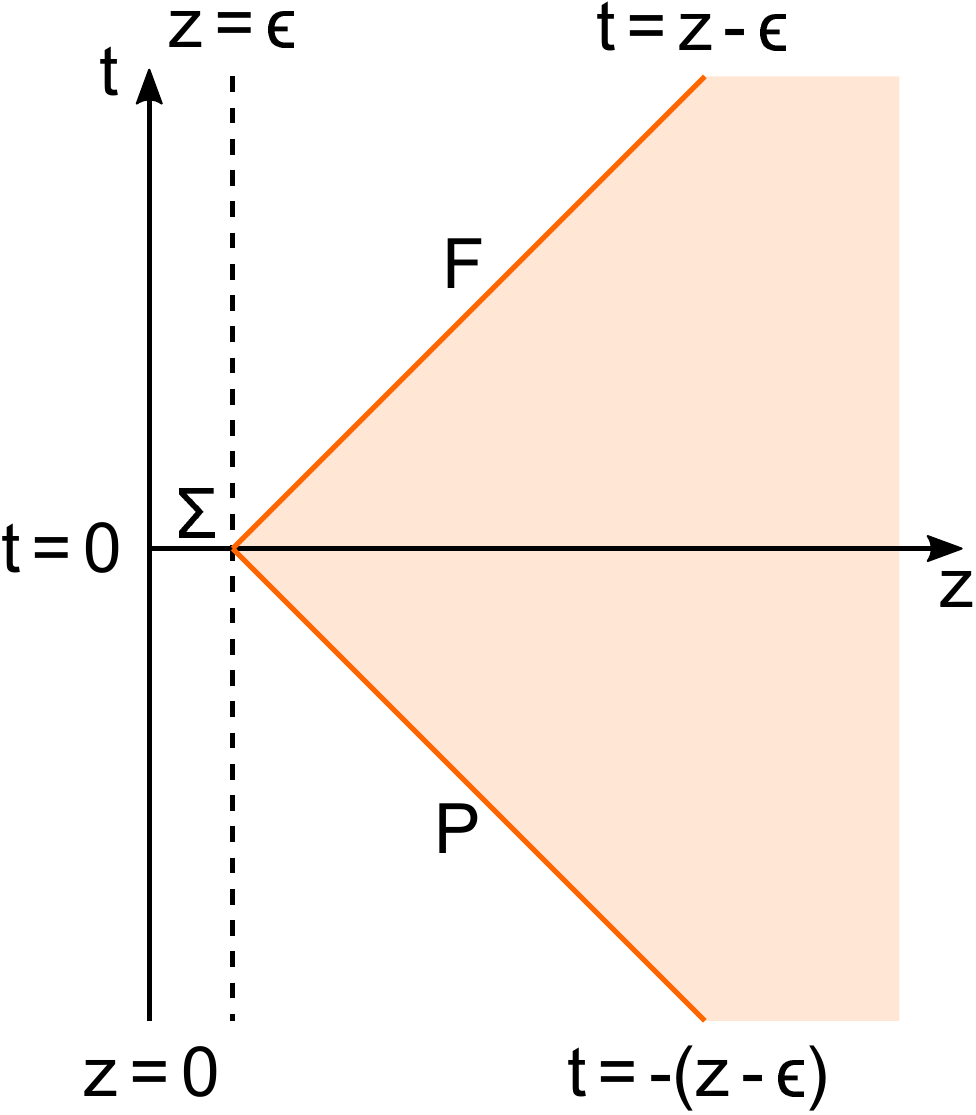}
\caption{The Wheeler-de Witt patch in Poincar\'e coordinates.} \label{WDW}
\end{figure}
With an affine parametrization along the null surfaces, 
\begin{equation} \label{afp}
\lambda = - \frac{\ell^2}{\alpha z} \mbox{ on }F, \quad \lambda = \frac{\ell^2}{\beta z} \mbox{ on }P,
\end{equation}
the action calculated according to \eqref{adsa} is 
\begin{equation}
S_{\cal V} = \frac{\ell^{d-1} V_x}{\epsilon^{d-1}}[ - 4 \ln( \epsilon/\ell) - 2 \ln(\alpha\beta) - \frac{1}{d-1} ].  
\end{equation}
We include the additional contribution \eqref{acorr}. The metric on $F$ has $\sqrt{\gamma} = \ell^{d-1}/z^{d-1}$, so the expansion is
\begin{equation}
\Theta = \frac{1}{\sqrt{\gamma}} \frac{\partial \sqrt{\gamma}}{\partial \lambda} = - \frac{1}{\sqrt{\gamma}} \alpha  \frac{z^2}{\ell^2}  \frac{\partial \sqrt{\gamma}}{\partial z} = (d-1) \alpha \frac{z}{\ell^2}. 
\end{equation}
The surface term is
\begin{eqnarray}
S_{\textnormal{F}} &=&  2 (d-1) \ell^{d-1} V_x \int_\epsilon^\infty z^{-d} \ln (\alpha (d-1) z/\ell) \mathop{dz} \nonumber \\  &=&  2\frac{ \ell^{d-1}}{\epsilon^{d-1}} V_x \left( \ln( \alpha (d-1) \epsilon/\ell) + \frac{1}{d-1} \right),  \nonumber 
\end{eqnarray}
and similarly $S_{\textnormal{P}} = 2\frac{ \ell^{d-1}}{\epsilon^{d-1}} V_x \left( \ln(\beta (d-1) \epsilon/\ell) + \frac{1}{d-1} \right)$, so 
\begin{equation} \label{adsaf}
S = S_{\cal V} + \Delta S = S_{\textnormal{Vol}} + S_\Sigma + S_{\textnormal{F}} + S_{\textnormal{P}} = 4  \frac{ \ell^{d-1}}{\epsilon^{d-1}} V_x \ln(d-1).  
\end{equation}
Taking the proposal \eqref{ca} for the complexity, this gives
\begin{equation} 
{\mathcal C_{\textnormal{A}}} = \frac{ 4 \ell^{d-1} V_x}{\pi \epsilon^{d-1}}  \ln(d-1).  
\end{equation}
This has the same divergence structure as in \eqref{pvol}, but with a different coefficient. Subleading divergences in both calculations can be expressed in terms of local geometric properties of the boundary in both cases, but the relative coefficients differ.   

\section{AdS soliton}
\label{solitons}

If we consider a field theory defined on a flat torus, with periodic boundary conditions for the fermions (preserving supersymmetry), the holographic dual of the ground state is the pure AdS solution in Poincare coordinates considered above. If however we take antiperiodic boundary conditions for the fermions on one or more directions, while the pure AdS solution is still a solution, it no longer corresponds to the ground state in the field theory. The holographic dual of the ground state is instead the AdS soliton \cite{Horowitz:1998ha},
\begin{equation} \label{soliton}
ds^2 = \frac{r^2}{\ell^2} \left[ - dt^2 + \left( 1 - \frac{r_+^d}{r^d} \right) d\chi^2 + d\vec x^2  \right] + \left( 1 - \frac{r_+^d}{r^d} \right)^{-1} \frac{\ell^2}{r^2} dr^2, 
\end{equation}
where $\chi$ is the circle with antiperiodic boundary conditions (or if there is more than one such circle, the one with the smallest period). We take a $d$-dimensional boundary, so there are $d-2$ coordinates $\vec x$. Imposing smoothness at $r=r_+$ relates the parameter $r_+$ to the periodicity of $\chi$, $\Delta \chi = \frac{4 \pi \ell^2}{d r_+}$. This solution has a negative boundary energy, 
\begin{equation} 
E = - \frac{r_+^d \Delta \chi V_x}{\ell^{d+1}} = - \frac{V_x \ell^{d-1} (4 \pi)^d}{d^d \Delta \chi^{d-1} }. 
\end{equation}
This can be understood as a Casimir energy for the ground state due to the periodicity of $\chi$. Because of the antiperiodic boundary conditions for the fermions, the Casimir energies of bosons and fermions fail to cancel. 

Note that the dependence on $r_+$ can be converted into an overall scale by a change of coordinates: if we set
\begin{equation} 
r =r_+ \tilde r, \quad t = \frac{\tilde t}{r_+}, \quad  \chi = \frac{\tilde \chi}{r_+}, \quad  x^i = \frac{\tilde x^i}{r_+}, 
\end{equation}
the metric becomes 
\begin{equation} \label{tsoliton}
ds^2 = \frac{\tilde r^2}{\ell^2} \left[ - d\tilde t^2 + \left( 1 - \frac{1}{\tilde r^d} \right) d\tilde \chi^2 + d\vec \tilde x^2  \right] + \left( 1 - \frac{1}{\tilde r^d} \right)^{-1} \frac{\ell^2}{\tilde r^2} d\tilde r^2. 
\end{equation}
Let us write for later convenience 
\begin{equation} 
f(r) = 1 - \frac{r_+^d}{r^d} = 1 - \frac{1}{\tilde r^d}. 
\end{equation}

It is interesting to consider the complexity of the ground state for the field theory with these boundary conditions, and specifically its dependence on the size of the $\chi$ circle. For the CV conjecture, the maximum volume calculation is easily carried out. Because of the time-independence of the metric \eqref{soliton}, the maximum volume slice will lie at constant $t$, so the volume is simply 
\begin{equation} \label{volsol}
V(B) = \int \mathop{dr}\mathop{d^{d-2}x} d\chi \sqrt{h} = V_x \Delta \chi  \int_{r_+}^{r_{max}} \frac{r^{d-2}}{\ell^{d-2}} = \frac{V_x \Delta \chi}{d-1} \frac{r_{max}^{d-1} - r_+^{d-1}}{\ell^{d-2}},
\end{equation}
where we introduce a UV cutoff at $r=r_{max}$. This gives us a complexity 
\begin{equation} \label{cvol}
{\cal C}_V = \frac{8 V_x \Delta \chi}{\pi} \frac{r_{max}^{d-1} - r_+^{d-1}}{\ell^{d-1}}.
\end{equation}
The first term is the same UV divergence we saw in  the pure AdS solution in \eqref{pvol}. If we were to take the difference, defining a `complexity of formation' as in \cite{Chapman:2016hwi}, there is a finite negative difference;  changing the boundary conditions has lowered the complexity. Put another way, the pure AdS solution, which corresponds to some excited state with these boundary conditions, has higher complexity than the ground state.  This seems a plausible result; adding excitations might be expected to generically increase the complexity of the state.

If we were to take $r_+ \to r_{max}$, the complexity would go to zero. As this limit corresponds to the proper size of the $\chi$ circle at the UV cutoff scale vanishing, this seems physically reasonable.  Note that the complexity vanishes linearly in $r_{max} - r_+$ in this case. 

\section{Holographic action calculations}
\label{action}

We now turn to the calculation of the complexity using the CA conjecture, calculating  the action of the Wheeler-de Witt patch for the AdS soliton. We will find that the action of the Wheeler-de Witt patch initially increases with $r_+$, although it does ultimately go to zero as $r_+ \to r_{max}$ as well. 

The calculation of the action is quite similar to the calculation in the bubbles with de Sitter boundaries in our previous work \cite{Reynolds:2017lwq}, although somewhat simpler. Indeed,  in the limit of large $r_+$, the  Wheeler-de Witt patch in those bubble solutions approaches the Wheeler-de Witt patch in the AdS soliton. 

The action will have an overall scaling as $r_+^{d-1}$, which is evident if we perform the calculation in the rescaled coordinates of \eqref{tsoliton}. If the original coordinates have a UV cutoff at $r = r_{max}$, then in the tilded coordinates $\tilde r \in (1, r_{max}/r_+)$, so the result of the action integrals will be some function of $ r_{max}/r_+$, times the coordinate volume in the spatial directions 
\begin{equation}
S = \tilde V_x  \Delta \tilde \chi I( r_{max}/r_+). 
\end{equation}
If we rewrite the spatial volumes in terms of the original coordinates, we get an overall factor of $r_+^{d-1}$. Thus 
\begin{equation} \label{rpscal}
S =  V_x \Delta \chi  r_+^{d-1} I( r_{max}/r_+). 
\end{equation}
If we take the UV cutoff large at fixed $r_+$, there will be a power series expansion in powers of $r_{max}/r_+$. From the results of \cite{Carmi:2016wjl}, we know the divergent terms in this expansion will be determined by the local geometric invariants of the boundary. For the flat boundary we are considering, the only non-zero term is the leading divergence, proportional to the volume, which agrees with the result in the pure AdS case. Thus, in the large $r_{max}$ limit, the action looks like 
\begin{equation}
S =  4 V_x  \Delta \chi \ln(d-1)  \frac{r_+^{d-1}}{\ell^{d-1}}   \left[ \left( \frac{r_{max}}{r_+} \right)^{d-1} + I_0 + \ldots \right] =   \frac{4  V_x  \Delta \chi}{\ell^{d-1}} \ln(d-1)  ( r_{max}^{d-1} + I_0 r_+^{d-1} + \ldots),  
\end{equation}
where the dots denote terms which vanish in the limit of large $r_{max}$. Thus, as in the volume calculation above, there is a finite difference between the complexity with antiperiodic and periodic boundary conditions, determined by the numerical parameter $I_0$. We will calculate the action in detail to determine $I_0$; from our previous work on the de Sitter case \cite{Reynolds:2017lwq}, we expect it to be positive, in contrast to the CV calculation. 

In our numerical calculation of the action, we add to \eqref{adsaf} a local integral over $\Sigma$ constructed to cancel the leading divergence;
\begin{equation}
S_{ct}  =  -4 \ln(d-1)  \int_\Sigma \sqrt{h} dS = -  \frac{4  V_x  \Delta \chi}{\ell^{d-1}} \ln(d-1) r_{max}^{d-1} \sqrt{ 1 - \frac{r_+^d}{r_{max}^{d}} }.  
\end{equation}
Thus, the action $S' = S + S_{ct}$ has $I_0$ as its leading contribution at large $r_{max}/r_+$. This change in the action is a part of the remaining ambiguity unfixed by the prescription of \cite{Lehner:2016vdi}. The interpretation of this kind of renormalization in terms of the complexity is unclear, but it is convenient for the numerics, and since the counterterm is a known function, one can remove it at the end of the calculation if desired. Note that the subleading contribution in $S_{ct}$ is of order $r_+^d/r_{max}$, so adding this term does not affect the finite contribution $I_0$.  

We now turn to the details of the calculation of the action. In the metric \eqref{soliton}, if we take a slice of the boundary at $t = 0$, the null boundaries of the Wheeler-de Witt patch are given by 
\begin{equation}
t(r) = \pm \ell^2 \int_r^{r_{max}} \frac{dr'}{r^{'2} \sqrt{f(r')}} ,
\end{equation}
or in terms of the tilded coordinates,
\begin{equation}
\tilde t =  \pm \ell^2 \int_{\tilde r}^{r_{max}/r_+} \frac{d\tilde r'}{\tilde r^{'2} \sqrt{f(\tilde r')} }  .
\end{equation}
 The volume integral is 
\begin{equation} \label{svol}
\begin{split}
S_{\textnormal{Vol}} &= - \frac{2d V_{\vec x} \Delta \chi}{\ell^2} \int_{r_+}^{r_{\textnormal{max}}} \mathop{dr} \frac{r^{d-1}}{\ell^{d-1}} 2 t(r) \\
&= - \frac{2d V_{\vec x} \Delta \chi r_+^{d-1}} {\ell^{d+1}} \int_{1}^{r_{\textnormal{max}}/r}  \mathop{d\tilde r} \tilde r^{d-1} 2 \tilde t(\tilde r). 
\end{split}
\end{equation}
If we write
\begin{equation}
F(r) =  \int_r^{r_{max}} \frac{dr'}{r'^2}  \left( 1 - \frac{r_+^d}{r^{'d}} \right)^{-1/2} = \int_{\tilde r}^{r_{max}/r_+} \frac{dr'}{r'^2}  \left( 1 - \frac{1}{r^{'d}} \right)^{-1/2}, 
\end{equation}
we can write this as 
\begin{equation}
S_{\textnormal{Vol}} =  - \frac{4d V_{\vec x} \Delta \chi r_+^{d-1}}{\ell^{d-1}} \int_{1}^{r_{\textnormal{max}}/r_+} \tilde r^{d-1} F \mathop{d \tilde r}.   
\end{equation}

The tangent to the null surface is 
\begin{equation}
k = \alpha ( \frac{\ell}{r^2} \mathop{\partial_t} - \frac{\sqrt{f(r)}}{\ell} \mathop{\partial_r}) 
\end{equation}
on $F$ and 
\begin{equation}
\bar k = \beta (  \frac{\ell}{r^2} \mathop{\partial_t} + \frac{\sqrt{f(r)}}{\ell} \mathop{\partial_r}) 
\end{equation}
on $P$, where $\alpha$, $\beta$ are some arbitrary positive constants. The corner term in the action is 
\begin{equation}
S_\Sigma = - 2 V_{\vec x} \Delta \chi r_{\textnormal{max}}^{d-1} \sqrt{f(r_{\textnormal{max}})}\ln\left( \frac{\alpha \beta}{r_{\textnormal{max}}^2} \right). 
\end{equation}

The expansions are
\begin{equation}
\Theta_F = - \alpha \frac{\sqrt{f}}{\ell} \frac{1}{\sqrt{\gamma}} \frac{\partial \sqrt{\gamma}}{\partial r} = - \alpha  \frac{\sqrt{f}}{\ell} \left( \frac{1}{2} \frac{f'}{f} + \frac{(d-1)}{r} \right) 
\end{equation}
and
\begin{equation}
\Theta_P = \beta \frac{\sqrt{f}}{\ell} \frac{1}{\sqrt{\gamma}} \frac{\partial \sqrt{\gamma}}{\partial r} = \beta  \frac{\sqrt{f}}{\ell} \left( \frac{1}{2} \frac{f'}{f} + \frac{(d-1)}{r}  \right) 
\end{equation}
so the surface integrals are
\begin{equation} \label{sf}
S_{\textnormal{F}} = 2 V_{\vec x} \Delta \chi \frac{r_+^{d-1}}{\ell^{d-1}}   \int_{1}^{r_{\textnormal{max}}/r_+} \sqrt{f} r^{d-1}  \left( \frac{1}{2} \frac{f'}{f} + \frac{(d-1)}{r} \right) \ln |\ell \Theta_F| \mathop{dr} 
\end{equation}
and
\begin{equation} \label{sp}
S_{\textnormal{P}} =2 V_{\vec x} \Delta \chi \frac{r_+^{d-1}}{\ell^{d-1}} \int_{1}^{r_{\textnormal{max}}/r_+} \sqrt{f} r^{d-1}  \left( \frac{1}{2} \frac{f'}{f} + \frac{(d-1)}{r}  \right) \ln |\ell \Theta_P| \mathop{dr} 
\end{equation}
So the total integral is 
\begin{eqnarray}
S' &=& \frac{2 V_{\vec x} \Delta \chi}{\ell^{d-1}}  \left[  - 2dr_+^{d-1}   \int_{1}^{r_{\textnormal{max}}/r_+} r^{d-1}  F \mathop{dr} - \sqrt{f(r_{\textnormal{max}})} r_{\textnormal{max}}^{d-1} \ln\left( \frac{\alpha \beta}{r_{\textnormal{max}}^2} \right) \right. \nonumber
\\ &&+ r_+^{d-1} \int_{1}^{r_{\textnormal{max}}/r_+} \sqrt{f} r^{d-1} \left( \frac{1}{2} \frac{f'}{f} + \frac{(d-1)}{r}  \right) \ln |\ell \Theta_F| \mathop{dr}  \nonumber
\\ &&  + r_+^{d-1}  \int_{1}^{r_{\textnormal{max}}/r_+} \sqrt{f} r^{d-1} \left( \frac{1}{2} \frac{f'}{f} + \frac{(d-1)}{r}  \right) \ln |\ell \Theta_P| \mathop{dr} \nonumber 
\\ && \left. -2 \ln(d-1) r_{max}^{d-1} \sqrt{f(r_{max})}  \right]. 
\end{eqnarray}

\begin{figure}
\centering 
\includegraphics[width=0.6\textwidth]{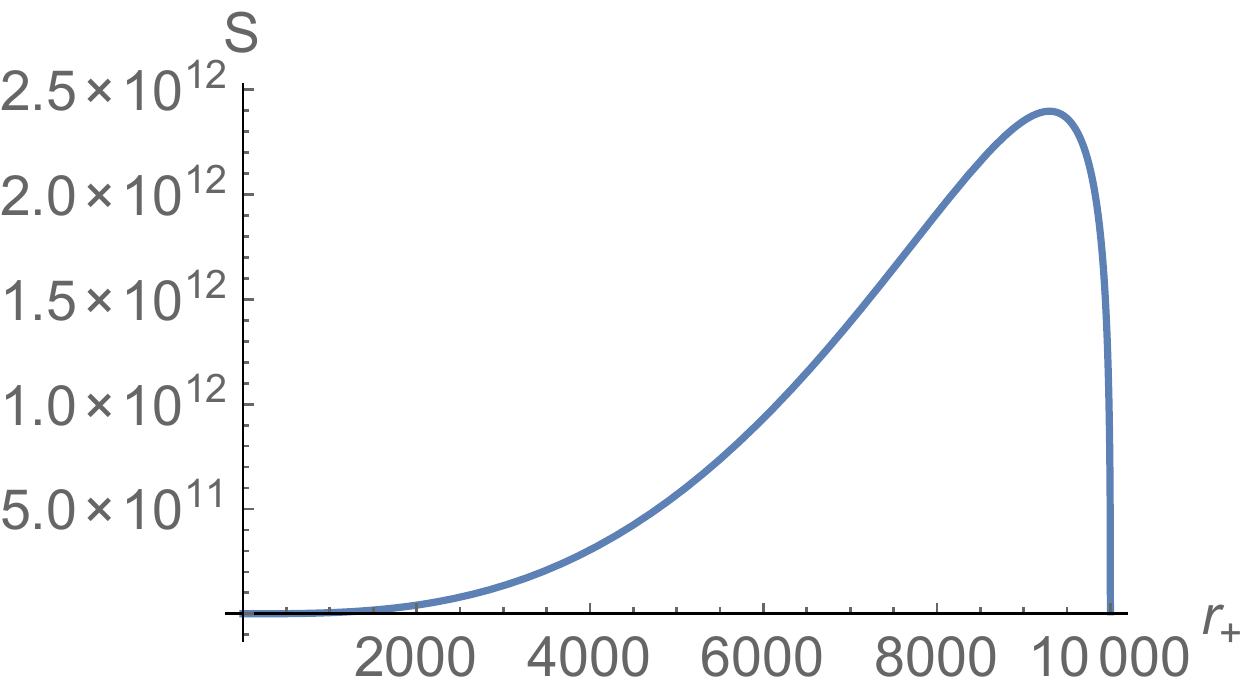}
\caption{We plot the action, omitting the overall factor of $2 V_{\vec x} \Delta \chi / \ell^{d-1}$, as a function of $r_+$ at fixed $r_{max} = 10000$. We see that it initially increases with $r_+$, but eventually decreases to zero as $r_+ \to r_{max}$. The initial increase scales as $r_+^{d-1}$, as indicated by the general scaling argument, with $I_0$ found to be approximately 1.27.} \label{actionplot}
\end{figure}

\begin{figure}
\centering 
\includegraphics[width=0.6\textwidth]{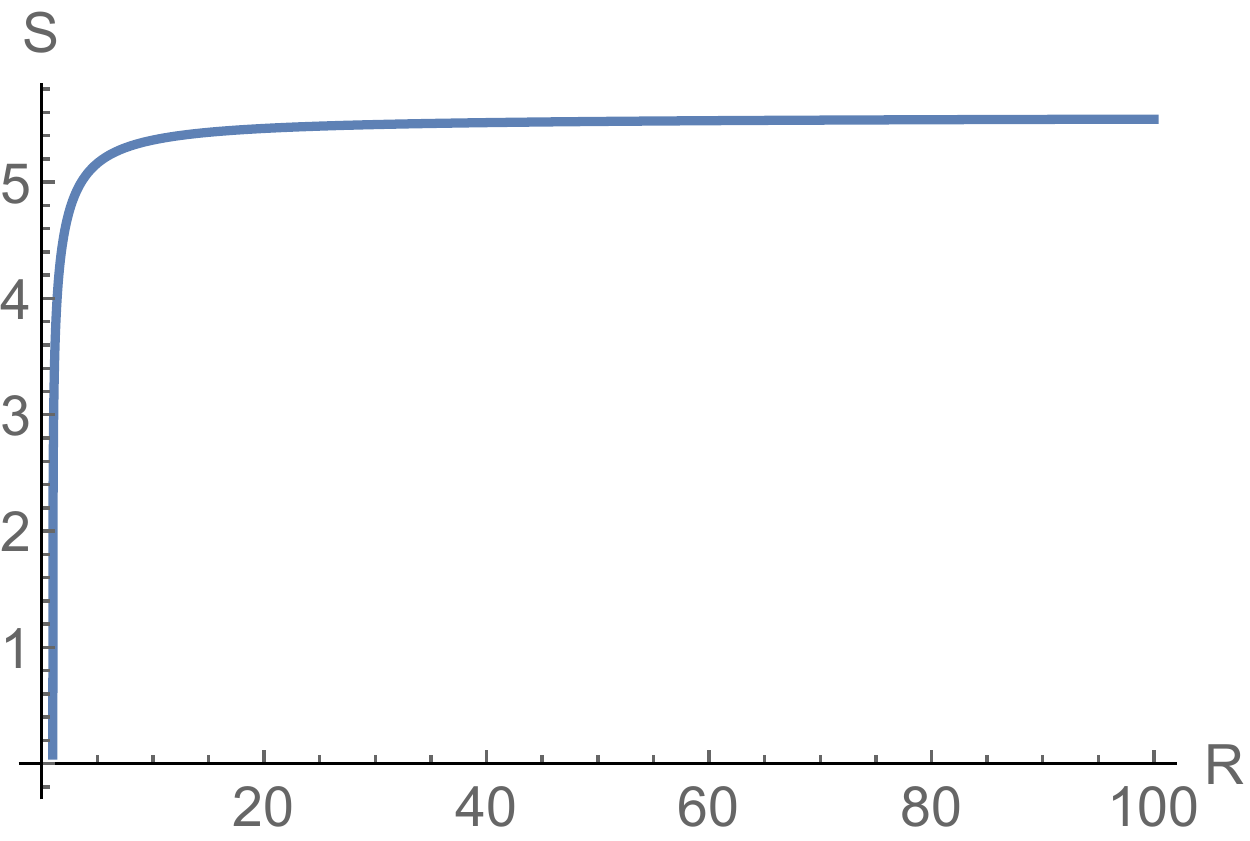}
\caption{We plot the action, omitting the overall factor of $2 V_{\vec x} \Delta \chi / \ell^{d-1}$, as a function of $r_{max}$ at fixed $r_+=1$. We see that it is a monotonically increasing function of $r_{max}$, which seems physically reasonable behaviour for the complexity.} \label{actionplot2}
\end{figure}

This integral is a function of $r_{max}$ and $r_+$, which is homogeneous of degree $d-1$. It is straightforward to evaluate these expressions numerically for fixed values of the parameters. In figure \ref{actionplot}, we plot the action as a function of $r_+$ at fixed $r_{max}$; in figure \ref{actionplot2}, we plot the action as a function of $r_{max}$ at fixed $r_+$. 

We see that the action initially increases at small $r_+$, indicating that $I_0$ is positive (approximately 1.27). This is qualitatively different from the behaviour of the volume \eqref{volsol}.  The increase comes basically from the negative volume contribution; increasing $r_+$ decreases the volume of the Wheeler-de Witt patch, and the volume contribution to the action is negative. 

In the CA calculation, the complexity for antiperiodic boundary conditions is higher than for periodic boundary conditions. Equally, the complexity for the excited state represented by the pure AdS solution with antiperiodic boundary conditions is lower than that of the ground state. 

We see that numerically the action goes to zero as $r_+ \to r_{max}$. It is interesting to compare the approach to zero in this regime to the volume calculation \eqref{volsol}. Suppose $r_{max} - r_+ \ll r_+$, and define $\epsilon = r_{max}/r_+ -1$.  Set $ r = r_+ (1 + \epsilon z)$, so $z \in (0,1)$. Then 
\begin{equation} 
f(r) \approx (r-r_+) f'(r_+) \approx d \epsilon z, 
\end{equation}
and $F(r) \sim \int dr/ \sqrt{f(r)}$ scales as $\sqrt{\epsilon}$, so that the  volume contribution to the action scales as $\epsilon^{3/2}$. The contributions from $\Sigma$ scale as $\sqrt{f}$, that is as $\sqrt{\epsilon}$, but the slowest falloff comes from the expansion contributions on the null surfaces: 
\begin{equation} 
\Theta_{F,P} \sim \frac{f'}{\sqrt{f}} \sim \frac{1}{\sqrt{\epsilon}},  
\end{equation}
so 
\begin{equation} 
S_{F,P}  \sim \int \frac{f'}{\sqrt{f}} \ln| \ell \Theta_{F,P}| dr  \sim \sqrt{\epsilon} \ln \epsilon.   
\end{equation}
Thus, the action goes to zero more slowly than the volume, as $\sqrt{r_{max} - r_+} \ln(r_{max} - r_+)$.

\section{Lattice calculations}
\label{lattice}

The holographic complexity calculations gave qualitatively different answers for the two calculation methods. It is interesting to understand what notions of complexity we can identify in the field theory that could reproduce these behaviours.  Since the AdS soliton is distinguished by the boundary conditions for fermions, we want to consider a fermionic theory. We consider free fermions on a rectangular lattice. We will consider both the case where the fermions have conventional periodic boundary conditions on all the spatial directions, and a lattice with antiperiodic boundary conditions on one spatial direction and periodic boundary conditions in the remaining directions, and study the difference between the complexity with the antiperiodic boundary conditions and the complexity with the periodic boundary conditions as a function of the size of the spatial direction with the changing boundary conditions. 

Complexity for fermionic field theories was previously considered in \cite{Jordan:2014tma}. Our analysis will also draw inspiration from the recent study of scalar field theories \cite{Jefferson:2017sdb}, where connections to the holographic calculation were also considered.   

\subsection{Free fermion theory}

We will first review the details that we need of the lattice fermion theory. We consider a theory of a single free fermion $\psi(\vec x)$ on a spatial lattice.  We will discuss explicitly lattices in two and three dimensional spacetimes. The generalisation to higher dimensions has some additional technical complications, so we leave it to the appendix. 

\subsubsection{Two dimensions}

The simplest case is two dimensions. Then the lattice has a single spatial direction; we can take either periodic or antiperiodic boundary conditions on this direction. The absence of additional spatial directions with periodic boundary conditions makes this case rather special.\footnote{From the holographic perspective, with a two-dimensional boundary the AdS soliton is actually global AdS$_3$, and the geometry in the bulk does not change as we vary $\Delta \chi$.}  We have $N$ lattice sites, $x_i = i a$, $i = 0, \ldots, N$, with $x_N = x_0$, and the boundary condition is $\psi(x_N) = \pm \psi(x_0)$ for the periodic (antiperiodic) case respectively. The Hamiltonian of the free fermion theory is 
\begin{multline}
H = a \sum_{i = 0}^{N-1} \biggl[  m \bar \psi(x_i) \psi(x_i) - i \bar \psi(x_i) \gamma^1 \frac{(\psi(x_i+a)-\psi(x_i-a) )}{2a}\\
 - r \bar \psi(x_i) \frac{( \psi(x_i + a) - 2 \psi(x_i) + \psi(x_i-a))}{2a} \biggr],   
\end{multline}
where the last term is the Wilson term, used to prevent fermion doubling \cite{Wilson:1974sk}. The fermion $\psi$ has two components, we define $\bar \psi = \psi^\dagger \gamma^0$, and we work with the gamma matrix representation\footnote{Note that our Clifford algebra conventions correspond to taking the lattice theory's spacetime metric to be $ds^2 = dt^2 - dx^2$, the opposite sign convention to our holographic discussion. We have adopted this convention for consistency with standard references.}  
\begin{equation} 
\gamma^0 = \left[ \begin{array}{cc} 0 & -i \\ i & 0 \end{array} \right], \quad \gamma^1 = \left[ \begin{array}{cc} 0 & -i \\ -i & 0 \end{array} \right].
\end{equation}
We will primarily work in momentum space, writing $\psi(x_j) = \frac{1}{\sqrt{N}} \sum_{i=0}^{N-1} e^{-i p_i x_j} \psi(p_i)$, where the momentum lives in the dual lattice or Brillouin zone; for periodic boundary conditions,
\begin{equation} \label{perl}
p_i = \frac{2 \pi}{Na} i, \quad i \in \mathbb{Z}_N,
\end{equation}
while for antiperiodic boundary conditions
\begin{equation} \label{aperl}
p_i = \frac{2 \pi}{Na} (i+ \frac{1}{2}). 
\end{equation}
The Hilbert space can be written as a tensor product of the Hilbert space $\mathcal H_{p}$ acted on by the fermionic operators $\psi(p)$ at each momentum. The Hamiltonian in momentum space is 
\begin{equation} 
H =  a \sum_{p \in \Omega} \left[  m \bar \psi(p) \psi(p)  +\frac{\sin p a}{a} \bar \psi(p) \gamma^1 \psi(p)  + \frac{2r}{a} \sin^2 \left(\frac{p a}{2} \right) \bar \psi(p) \psi(p) \right],   
\end{equation}
where $\Omega$ is the lattice of momentum values in \eqref{perl} or \eqref{aperl} depending on the boundary conditions.  The term in the Hamiltonian at a given momentum has eigenspinors\footnote{Usually the convention for negative energy states associates them with momentum $-p$; the convention here will be more convenient for our calculations.} 
\begin{equation}
u = \frac{1}{\sqrt{2E}} \left[ \begin{array}{c} \sqrt{E -  P} \\ i \sqrt{E + P} \end{array}  \right], \quad v = \frac{1}{\sqrt{2E}} \left[  \begin{array}{c} \sqrt{E +  P} \\ -i \sqrt{E - P} \end{array}  \right],
\end{equation}
with eigenvalues $\pm E$, where 
\begin{equation}
P = \frac{\sin pa}{a}, \quad M =  m + \frac{2r}{a} \sin^2 \left(\frac{p a}{2} \right), \quad E = \sqrt{M^2 + P^2 }. 
\end{equation}
Thus, the fermion can be written in terms of ladder operators as 
\begin{equation}
\psi(p) = u(p) a(p) + v(p) b^\dagger(p), 
\end{equation}
and the ground state is the state annihilated by $a(p_i), b(p_i)$ for all $p_i$; it is the tensor product of the ground state in each $\mathcal H_{p}$. 

\subsubsection{Three dimensions}

For the three dimensional case, we can take a representation of the Clifford algebra where we enlarge the previous representation by adding 
\begin{equation}
\gamma^2 = \left[ \begin{array}{cc} i & 0 \\ 0 & -i \end{array} \right].
\end{equation}
The fermions still have two components. We will take the direction with variable boundary conditions to be the $x$ direction. The spatial lattice has $N_x$ sites in the $x$ direction with lattice spacing $a_x$, and $N_y$ sites in the $y$ direction with lattice spacing $a_y$. The momentum vector then lives in a lattice
\begin{equation} 
\vec p = (\frac{2 \pi}{N_x a_x} i, \frac{2 \pi}{N_y a_y} j) 
\end{equation}
for periodic boundary conditions, and 
\begin{equation} 
\vec p = ( \frac{2 \pi}{N_x a_x} (i+ \frac{1}{2}), \frac{2 \pi}{N_y a_y} j)
\end{equation}
for antiperiodic boundary conditions, where in both cases $i \in \mathbb Z_{N_x}$, $j \in \mathbb Z_{N_y}$. The Hilbert space is a tensor product of spaces $\mathcal H_{\vec p}$ associated with each lattice site. The Hamiltonian is 
\begin{eqnarray} 
H &=& a_x a_y \sum_{\vec p \in \Omega} \left[  m \bar \psi(\vec p) \psi(\vec p)  +\frac{\sin p_x a_x}{a_x} \bar \psi(\vec p) \gamma^1 \psi(\vec p) +\frac{\sin p_y a_y}{a_y} \bar \psi(\vec p) \gamma^2 \psi(\vec p)  \right. \\ && \left. + 2r  \left[ a_x^{-1} \sin^2 \left(\frac{p_x a_x}{2} \right) + a_y^{-1} \sin^2 \left(\frac{p_y a_y}{2}\right) \right] \bar \psi(\vec p) \psi( \vec p) \right], \nonumber
\end{eqnarray}
where $\Omega$ is the relevant momentum lattice. The eigenspinors at a given momentum are
\begin{equation}
u = \frac{1}{\sqrt{2E}} \left[ \begin{array}{c} \sqrt{E -  P_x} \\ i e^{i \beta_y} \sqrt{E + P_x} \end{array}  \right], \quad v = \frac{1}{\sqrt{2E}} \left[  \begin{array}{c} \sqrt{E +  P_x} \\ -i e^{i \beta_y} \sqrt{E - P_x} \end{array}  \right],
\end{equation}
with eigenvalues $\pm E$, where 
\begin{equation}
P_i = \frac{\sin p_i a_i}{a_i},\quad  M = m +  2r \sum_i a_i^{-1} \sin^2 \left(\frac{p_i a_i}{2} \right), \quad E = \sqrt{ M^2 + \vec P^2  }
\end{equation}
and
\begin{equation}
e^{i \beta_y} =\frac{ M+ i P_y }{\sqrt{M^2 + P_y^2  }}.
\end{equation}
Thus, the fermion can again be written as 
\begin{equation}
\psi(\vec p) = u(\vec p) a(\vec p) + v(\vec p) b^\dagger(\vec p), 
\end{equation}
and the ground state is the state annihilated by $a(\vec p), b(\vec p)$ for all $\vec p$; it is the tensor product of the ground state in each $\mathcal H_{\vec p}$.

\subsection{Complexity}

We wish to evaluate the complexity of the ground state in the free fermionic theories reviewed in the previous subsection. There are two key choices we need to make: we need to choose a reference state, and we need to define a measure of the complexity of the transformation from the reference state to the physical ground state. 

\subsubsection{Reference state}

In \cite{Jordan:2014tma}, the reference state is taken to be the ground state of the fiducial Hamiltonian 
\begin{equation} 
H_0 = a_x a_y \sum_{\vec x}  m_0 \bar \psi(\vec x) \psi(\vec x) = a_x a_y  \sum_{\vec p \in \Omega} m_0 \bar \psi(\vec p) \psi(\vec p),
\end{equation}
where the kinetic and Wilson terms are removed from the physical Hamiltonian. This Hamiltonian could also be viewed as a high-mass limit of our original Hamiltonian, where the momentum dependence becomes negligible.  This is a useful choice as the resulting reference state is a tensor product state in the position space representation and in the momentum space representation, so both the reference and target states are tensor products in the momentum space representation.  A similar choice was made in the scalar case in \cite{Jefferson:2017sdb}, where the reference state was taken to be a fixed Gaussian at each spatial lattice site; the tensor product of these Gaussian states in the spatial basis is also a tensor product of Gaussian states in the momentum basis. 

For the two and three-dimensional cases, the eigenspinors of this Hamiltonian are simply 
\begin{equation}
u_0 = \frac{1}{\sqrt{2}} \left[ \begin{array}{c} 1 \\ i  \end{array}  \right], \quad v_0 = \frac{1}{\sqrt{2}} \left[  \begin{array}{c} 1 \\ -i  \end{array}  \right].
\end{equation}
We can easily see that these are the high mass or low-momentum limit of the eigenspinors of the physical Hamiltonian found in the previous subsection. We write the spinor operator as 
\begin{equation}
\psi(\vec p) = u_0 a_0(\vec p) + v_0 b_0^\dagger(\vec p), 
\end{equation}
and we take the reference state to be the state annihilated by all the $a_0(\vec p), b_0(\vec p)$ for all $\vec p$. 

The physical creation and annihilation operators can be related to $a_0$ and $b_0$ by making use of the orthonormality of our eigenspinors, taking inner products in the spinor indices. In the two and three-dimensional cases,  
\begin{equation}
a(p) = u^\dagger(p) \psi(\vec p) = u^\dagger u_0 a_0(\vec p) + u^\dagger v_0 b_0^\dagger(\vec p), 
\end{equation}
\begin{equation}
b^\dagger(p) = v^\dagger(p) \psi(\vec p) = v^\dagger u_0 a_0(\vec p) + v^\dagger v_0 b_0^\dagger(\vec p).
\end{equation}
The key point is that this is a Bogoliubov-style transformation, mixing creation and annihilation operators, so the vacuum state with respect to $a(p)$, $b(p)$ will be a mixture of particle states with respect to $a_0(p)$, $b_0(p)$. For a given momentum, 
\begin{equation}
| 00 \rangle_{phys} = u^\dagger u_0 | 00 \rangle_0 - u^\dagger v_0 | 11 \rangle_0. 
\end{equation}
We note that this is a mixture of states with even numbers of fermions in the reference basis, so the transformation between the reference ground state and the physical ground state involves only fermion bilinears, so we do not need to worry about the difficulties of simulating fermionic gates reviewed for example in \cite{Jordan:2014tma}.  

For two dimensions, this gives 
\begin{eqnarray} \label{2dt}
| 00 \rangle_{phys} &=& \frac{1}{2 \sqrt{E}} \left[ (\sqrt{E+P} + \sqrt{E-P}) | 00 \rangle_0 - (\sqrt{E+P} - \sqrt{E-P}) | 11 \rangle_0 \right] \\
&=&  \sqrt{\frac{E+M}{2E}}    | 00 \rangle_0  - \sqrt{\frac{E-M}{2E}}  | 11 \rangle_0 . \nonumber
\end{eqnarray}
For three dimensions, we have 
\begin{equation} \label{3dt}
| 00 \rangle_{phys} = \frac{1}{2 \sqrt{E}} \left[ (e^{-i \beta_y} \sqrt{E+P_x} + \sqrt{E-P_x}) | 00 \rangle_0 - ( e^{-i \beta_y} \sqrt{E+P_x} - \sqrt{E-P_x}) | 11 \rangle_0 \right].  
\end{equation}
By changing the phase of the physical ground state, we can simplify this to 
\begin{equation} \label{3dt2}
| 00 \rangle_{phys} =  \sqrt{\frac{E+M}{2E}}  | 00 \rangle_0 - e^{i \phi_2} \sqrt{\frac{E-M}{2E}} |11 \rangle_0 ,
\end{equation}
where 
\begin{equation} 
e^{i \phi_2} = \frac{P_x - i P_y}{P_x + i P_y}. 
\end{equation}
Note that unlike the scalar case in \cite{Jefferson:2017sdb}, there is no dependence on the mass scale $m_0$ in the fiducial Hamiltonian. The reference ground state is the same, independent of which $H_0$ we choose.

\subsubsection{Complexity measure}

We want to compute the complexity of the least complex unitary operator $U$ such that the physical ground state $|\psi \rangle = U |\psi \rangle_0$, where $|\psi \rangle_0$ is the reference state.  Ideally, to respect the locality of the field theory, we would like to do this calculation taking as an elementary gate set some set of unitary operators which act on nearest neighbour sites in the spatial lattice. However, this calculation is extremely difficult, so following \cite{Jefferson:2017sdb}, we will make the simplifying assumption that we can take the elementary gate set to include unitary operators acting on the individual {\it momentum} sites in the momentum lattice. (Such operators can be built from a linear combination of operators acting on pairs of lattice sites in the spatial lattice, but we need to include arbitrary pairs of sites.)

Making this assumption allows us to exploit the special structure of our states: our ground state is the product of the ground state $| \vec 0 \rangle_{phys}$ in $\mathcal H_p$ at each momentum, and our reference state is the product of  $| \vec 0 \rangle_0$ in $\mathcal H_p$ at each momentum, so it is plausible that the least complex unitary will also have a tensor product decomposition, $U = \otimes_p U_p$. That is, we expect, as in \cite{Jefferson:2017sdb}, that the path of least complexity with such an elementary gate set will not involve introducing entanglement between different momenta at intermediate scales. Identifying the appropriate unitary $U$ then reduces to identifying an appropriate $U_p$ at each momentum. 

In the two and three-dimensional cases, the transformations (\ref{2dt},\ref{3dt}) from the reference state to the physical ground state at a given momentum can be implemented by a unitary transformation on the two-dimensional subspace of the Hilbert space spanned by $|00 \rangle_0$, $|11\rangle_0$. (We assume that considering more general transformations in $\mathcal H_p$ that take us out of this subspace will not reduce the complexity.) Such unitary transformations can be parametrized as
\begin{equation}
U_p = e^{i \alpha}  \left( \begin{array}{cc} e^{-i \phi_1} \cos \theta &  e^{-i \phi_2} \sin \theta \\ - e^{i \phi_2} \sin \theta & e^{- i\phi_1} \cos \theta \end{array} \right) .
\end{equation}
requiring that the unitary maps $|00 \rangle_0$ to $|00 \rangle_{phys}$ fixes the first column of $U_p$, giving three constraints on the parameters (since the target state is normalized, its form in terms of $|00\rangle_0$ and $|11\rangle_0$ involves three free parameters); that leaves one free parameter in $U_p$, which we need to minimize over.

As in \cite{Jefferson:2017sdb}, we will be inspired by the work of \cite{Nielsen1,Nielsen2,Nielsen3} to take a geodesic distance in a suitable metric in the space of unitaries as a proxy for the complexity. In the space at a given momentum, we will take the usual metric on $U(2)$,
\begin{equation} \label{2umet}
ds^2 = -\frac{1}{2} \mbox{tr} (dU U^{-1} dU U^{-1}) = d\alpha^2 + d \theta^2 + \cos^2 \theta d\phi_1^2 + \sin^2 \theta d\phi_2^2. 
\end{equation}
The remaining parameter in $U_p$ is determined by minimizing the distance from the identity in this metric.\footnote{This minimization to determine an appropriate $U_p$ becomes more difficult in higher dimensions; see the discussion for four dimensions in the appendix.} 

To calculate the overall complexity, we need to combine the complexities of the individual $U_p$ to obtain a complexity for $U$. We will simply sum up the complexities of each of the $U_p$: 
\begin{equation} \label{totc}
\mathcal C(U) = \sum_{p \in \Omega} \mathcal C(U_p).
\end{equation}
in the geometrical language of  \cite{Nielsen1}, this corresponds to taking an $F_1$  or ``Manhattan" metric, where the total distance is the sum of the distances along each of the basis directions. This is a natural choice for the calculation of complexity; it can be thought of as adding contributions from the different elementary gates acting on each $\mathcal H_p$. 

It is worth noting however that working in the Manhattan metric makes geometric analysis more challenging, which is why \cite{Nielsen2,Nielsen3} sought to replace it with a Riemannian metric with suitable cost factors on directions not corresponding to the elementary gate set. Notably, the Manhattan metric, unlike a Riemannian metric, depends on the choice of basis, so our choice to use gates acting on momentum subspaces in the Hilbert space rather than gates acting on pairs of position subspaces becomes significant. Even though the unitaries acting on a momentum subspace can be written as a linear combination of unitaries acting on position subspaces, taking one rather than the other as the basis in a Manhattan metric leads to a different formula for the complexity. 

The choice of a Manhattan metric is further supported by the results of \cite{Jefferson:2017sdb}, who found that  it reproduces the UV divergence structure of the holographic calculation. The same will be true in our fermionic calculation; the complexity defined by \eqref{totc} is roughly proportional to the number of lattice sites, which we can interpret as the volume in units of the UV cutoff, reproducing the divergence in the holographic calculation. Using a Riemannian metric to combine the metrics \eqref{2umet} would by contrast give a result scaling roughly as the square root of the number of sites. We will however not consider these divergent contributions further; our focus is on evaluating the finite difference between the complexities with periodic and antiperiodic boundary conditions,
\begin{equation} \label{deltac}
\Delta \mathcal C  = \mathcal C_{anti} - \mathcal C_{per}.
\end{equation}

We now implement this programme for the two- and three-dimensional cases.
In the two-dimensional case, the unitary of minimum distance which realises the transformation \eqref{2dt} is simply a rotation, 
\begin{equation}
U_p = \left( \begin{array}{cc} \cos \theta &  \sin \theta \\ -\sin \theta & \cos \theta \end{array} \right)   
\end{equation}
with
\begin{equation}
\cos \theta =  \sqrt{\frac{E+M}{2E}}. 
\end{equation}
The distance from the identity in the standard metric is simply $\theta$, so we take $\mathcal C(U_p) = \theta(p)$. This amounts to taking an infinitesimal rotation in this direction as an element of the elementary gate set. 

\begin{figure}
\centering 
\includegraphics[width=0.6\textwidth]{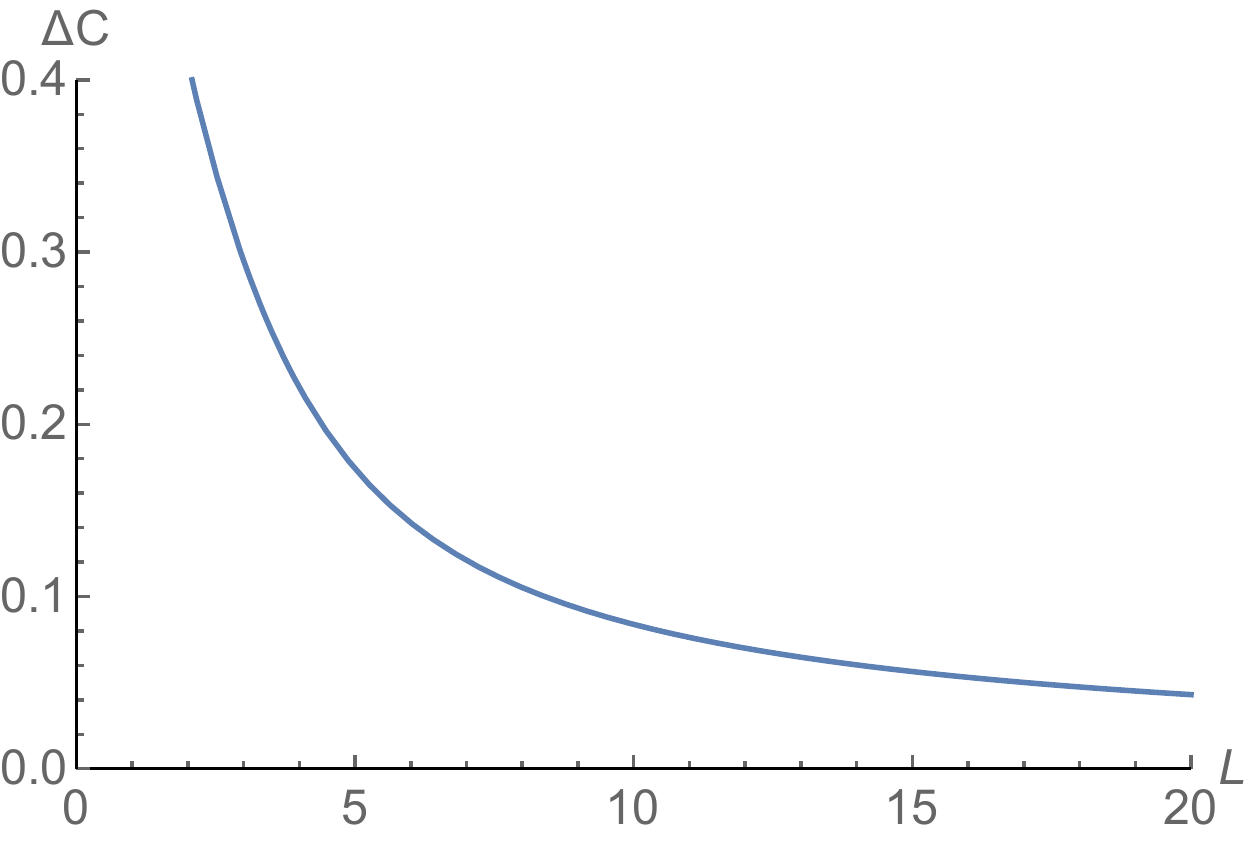}
\caption{We plot the difference $\Delta \mathcal C$ between the complexity of the ground states for a fermion with antiperiodic boundary conditions and a fermion with periodic boundary conditions on a one-dimensional spatial lattice, as a function of the size of the circle. We see that the difference is positive, and decreases as we increase the size of the circle.} \label{comp2d}
\end{figure}

We  sum over the contributions from the individual momenta, and take the difference between antiperiodic and periodic boundary conditions to calculate $\Delta \mathcal C$ in \eqref{deltac}. 
We plot this difference as a function of $L = a N$ in figure \ref{comp2d}.\footnote{For numerical convenience, we vary the length of the circle with a fixed number of sites, so we are varying both the size of the circle and the UV cutoff scale here. These results are thus not directly comparable to the holographic results, where we varied size of the circle at fixed UV cutoff scale. However, so long as the circle is large compared to the UV cutoff scale we would not expect $\Delta \mathcal C$ to depend on the UV cutoff scale, so this should not have a significant effect.} We see that the difference is positive, as for the holographic CA calculation, but unlike the holographic CV calculation. 

 We find that the difference decreases as the size of the circle increases. This is unlike the holographic calculation; there, the difference in complexity went as $\Delta \chi r_+^{d-1} \propto r_+^{d-2}$ (see \eqref{rpscal}), so in $d=2$, the difference in the holographic calculation is independent of the size of the circle. This is because in $d=2$ the AdS soliton is actually global AdS$_3$, and the finite part is the difference in volume or action between the $M=0$ BTZ black hole and global AdS$_3$, which is some finite constant. 

\begin{figure}
\centering 
\includegraphics[width=0.6\textwidth]{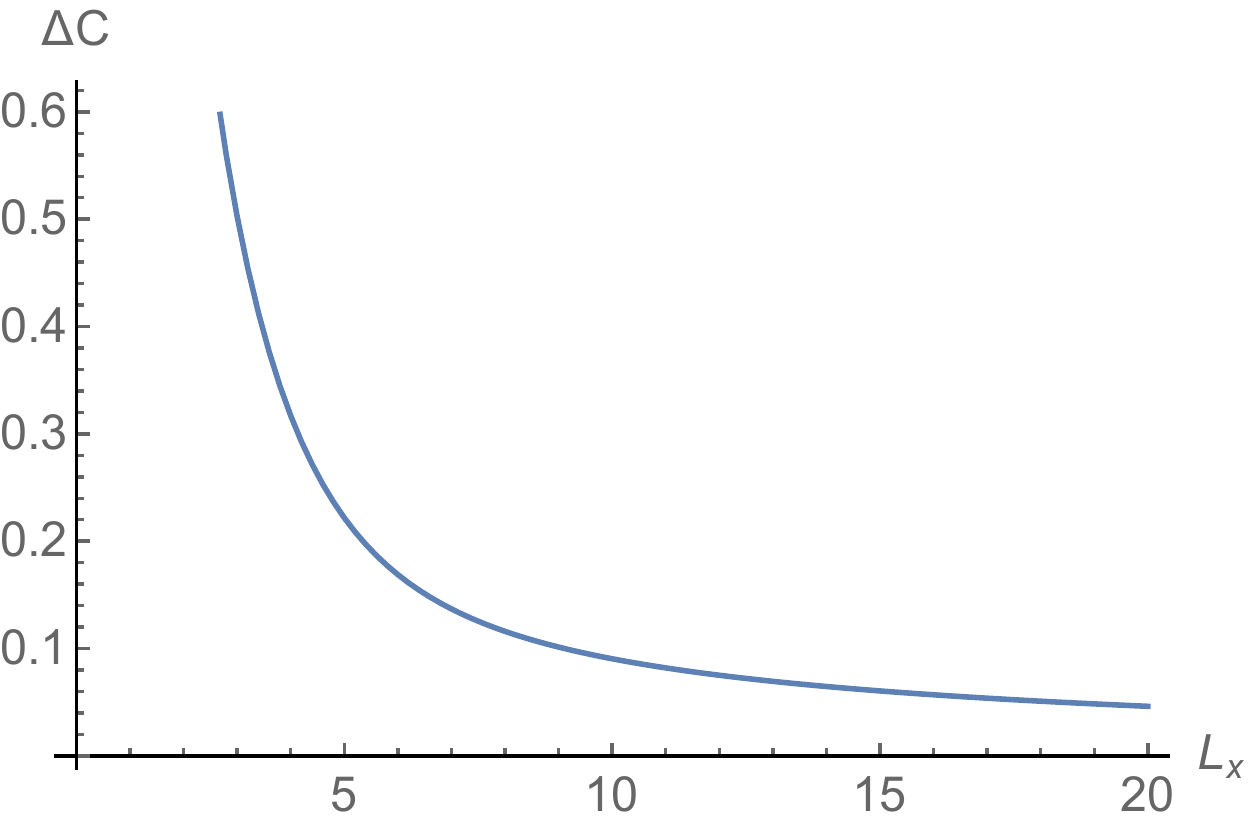}
\caption{We plot the difference $\Delta \mathcal C$ between the complexity of the ground states for a fermion with antiperiodic boundary conditions in one direction and a fermion with periodic boundary conditions in both directions on a two-dimensional spatial lattice, as a function of the size of the circle with the varying boundary conditions. We see that the difference is positive, and decreases as we increase the size of the circle.} \label{comp3d}
\end{figure}

In the three-dimensional case, the simplest unitary realising the transformation \eqref{3dt2} has $\alpha= 0, \phi_1 = 0$, 
\begin{equation} \label{ucond}
\cos \theta = \sqrt{\frac{E+M}{2E}} ,  \quad e^{i  \phi_2} =  \frac{P_x - i P_y}{P_x + i P_y}, 
\end{equation}
and the complexity is again $\mathcal C(U_p)  = \theta(p)$. We add up these contributions for each site in the momentum lattice, and take the difference between antiperiodic and periodic boundary conditions to calculate $\Delta \mathcal C$ in \eqref{deltac}. We plot this difference as a function of $L_x$ in figure \ref{comp3d}.\footnote{Note that the plot is again generated by varying the length of the circle with a fixed number of sites, so we are varying both the size of the circle and the UV cutoff scale.} We see that the difference is  again positive, and decreases as a function of the size of the circle. In this case this is also the behaviour expected holographically. Holographically, the complexity would fall off as $1/L_x$. The  numerical results for the lattice computation exhibit a faster falloff than the two-dimensional case, but it is not well fit by a simple power law. 

We have found that with our definition of the complexity, the complexity for antiperiodic boundary conditions is higher than for periodic boundary conditions. It seems surprising that the generic expectation that raising the energy increases the complexity is not borne out in this case. It is possible that the key ingredient in the increase in complexity in the antiperiodic case is the difference in the momentum lattices: in the antiperiodic case, the lowest momentum value is non-zero. Since the reference Hamiltonian is the zero-momentum limit of the physical Hamiltonian, this increase in the minimum momentum value may be responsible for the increase in complexity of the ground state relative to the reference state. 

In our calculation, we decomposed the unitary in terms of operators acting on different momentum subspaces of the Hilbert space. An important problem for the future is to study the decomposition in terms of operators acting on position subspaces and see if this modifies the results. Our use of the Manhattan metric makes this choice of basis particularly salient, and ultimately one would like to include appropriate penalty factors for non-local transformations in the position space decomposition.

\section{Discussion}
\label{disc}

We have studied the dependence of the complexity of the ground state of a field theory on a torus on the boundary conditions for fermions, both holographically and in a simple lattice model of a free fermion. We compared the results for  antiperiodic and periodic boundary conditions for the fermions. In the holographic calculation, the former corresponds to the AdS soliton, where the spacetime closes off at a radius $r_+$ determined by the size $\Delta \chi$ of the circle with antiperiodic boundary conditions, while the latter corresponds to a simple Poincare-AdS geometry. 

Without doing any calculations, we can argue that the holographic complexity for antiperiodic boundary conditions in the regime where $r_+$ is small compared to the UV cutoff $r_{max}$ will have the form
\begin{equation} 
\mathcal C_{anti} \propto V_{\vec x} \Delta \chi ( r_{max}^{d-1} + I_0 r_+^{d-1} + \ldots), 
\end{equation}
where $V_{\vec x}$ is the volume in the remaining spatial dimensions, $d$ is the spacetime dimension of the field theory, $I_0$ is a purely numerical coefficient, and the suppressed terms vanish in the limit as $r_{max} \to \infty$. Since the result for periodic boundary conditions is simply $\mathcal C_{per} \propto V_{\vec x} \Delta \chi  r_{max}^{d-1}$, the difference between periodic and antiperiodic boundary conditions is finite. 

By explicit calculation, we find that the result in the complexity volume (CV) calculation is $I_0 = -1$, while for the complexity action (CA) calculation $I_0$ is roughly 1.27.  This provides a qualitative distinction between these two bulk calculations. In the CV calculation, the change in boundary conditions reduces the complexity, while in the CA calculation it increases it. We initially thought a decrease was the more intuitive result, as the ground state with antiperiodic boundary conditions has lower energy than the one with periodic boundary conditions. However, the ground state is far from maximum complexity, so it is important to understand the overall effect of the change in boundary conditions. 

We investigated this in a simple lattice calculation for a free fermion, extending the work of  \cite{Jefferson:2017sdb} for bosons. The complexity of the fermion ground state is divergent, as for bosons, but the difference between the complexity for antiperiodic and periodic boundary conditions is finite. We find that this is positive, as in the holographic CA calculation. 

These results thus seem to support the holographic action calculations over the volume calculations. However, in the lattice calculation, there are a number of choices and approximations we need to make to render the calculation feasible, and improving this calculation is an important goal for future work. In particular, we would like to move from considering a basis of elementary operations that acts on the factors in a momentum space decomposition of the Hilbert space to one that acts in a position space decomposition, and ultimately to incorporate spatial locality into the calculation by penalizing operations that are not acting on nearest neighbour sites in position space.  

It would also be interesting to study the dependence of the complexity on other changes in the ground state, for example if we deform the field theory by relevant or marginal operators.

\section*{Acknowledgements}
AR is supported by an STFC studentship. SFR is supported in part by STFC under consolidated grants ST/L000407/1 and ST/P000371/1. We are grateful for helpful discussions with Nabil Iqbal, Ro Jefferson and Alex Maloney. 

\section*{Appendix: Lattice fermions in higher dimensions}

Here we discuss the lattice fermion theory in the four dimensional case in some detail, and comment on the extension to higher dimensions. In these cases the dimension of $\mathcal H_p$ is larger, and as a consequence the transformation between the reference state and the physical ground state is more involved. In four dimensions, we take the Dirac representation of the Clifford algebra, with 
\begin{equation}
\gamma^0 = \left[ \begin{array}{cc} 1 & 0 \\ 0 & -1 \end{array} \right], \quad \gamma^i = \left[ \begin{array}{cc} 0 & \sigma^i \\ - \sigma^i & 0 \end{array} \right],
\end{equation}
where each entry represents a $2\times 2$ matrix, and $\sigma^i$ are the Pauli matrices.  
The fermions then have four components. We will take the direction with variable boundary conditions to be the $x$ direction again. The spatial lattice has $N_x$ sites in the $x$ direction with lattice spacing $a_x$, $N_y$ sites in the $y$ direction with lattice spacing $a_y$, and $N_z$ sites in the $z$ direction with lattice spacing $a_z$. The momentum vector then lives in a lattice
\begin{equation} 
\vec p = (\frac{2 \pi}{N_x a_x} i, \frac{2 \pi}{N_y a_y} j, \frac{2 \pi}{N_z a_z} k ) 
\end{equation}
for periodic boundary conditions, and 
\begin{equation} 
\vec p = ( \frac{2 \pi}{N_x a_x} (i+ \frac{1}{2}), \frac{2 \pi}{N_y a_y} j, , \frac{2 \pi}{N_z a_z} k)
\end{equation}
for antiperiodic boundary conditions, where in both cases $i \in \mathbb Z_{N_x}$, $j \in \mathbb Z_{N_y}$, $k \in \mathbb Z_{N_z}$.  The Hilbert space is a tensor product of spaces $\mathcal H_{\vec p}$ associated with each lattice site. The Hamiltonian is 
\begin{equation}  \label{4dH}
H = a_x a_y a_z \sum_{\vec p \in \Omega} \left[  m \bar \psi(\vec p) \psi(\vec p)  + \sum_i \frac{\sin p_i a_i}{a_i} \bar \psi(\vec p) \gamma^i \psi(\vec p)   + 2r  \sum_i a_i^{-1} \sin^2 \left(\frac{p_i a}{2} \right) \bar \psi(\vec p) \psi( \vec p) \right], \nonumber
\end{equation}
where $\Omega$ is the relevant lattice. A convenient choice of eigenstates are 
\begin{equation}
 u^1 = \frac{1}{\sqrt{2E(E+M)}} \left[ \begin{array}{c} M+E \\ 0 \\ P_z \\ P_x + i P_y \end{array}  \right], \quad u^2 = \frac{1}{\sqrt{2E(E+M)}} \left[ \begin{array}{c} 0 \\ M+E \\ P_x - i P_y \\ - P_z \end{array}  \right], 
\end{equation}
with eigenvalue $E$, and
\begin{equation}
v^1 =  \frac{1}{\sqrt{2E(E+M)}} \left[ \begin{array}{c} -P_z \\ -P_x - i P_y \\ M+E \\ 0  \end{array}  \right]\quad
 v^2 =  \frac{1}{\sqrt{2E(E+M)}} \left[ \begin{array}{c} -P_x + i P_y \\ P_z \\0\\ M+E  \end{array}  \right],  \end{equation}
with eigenvalue $-E$, where 
\begin{equation}
P_i = \frac{\sin p_i a_i}{a_i}, \quad M^2 = m^2 +  2r  \sum_i a_i^{-1} \sin^2 \left(\frac{p_i a_i}{2} \right), \quad E = \sqrt{M^2 + \vec P^2}. 
\end{equation}
Writing the fermion as 
\begin{equation}
\psi(\vec p) = u^\alpha(\vec p) a^\alpha(\vec p) + v^\alpha(\vec p) b^{\alpha \dagger}(\vec p), 
\end{equation}
$\alpha = 1,2$,  the ground state is the state annihilated by $a^\alpha(\vec p), b^\alpha(\vec p)$ for all $\vec p$; it is the tensor product of the ground state in each $\mathcal H_{\vec p}$. 

Taking the reference Hamiltonian $H_0 = a_x a_y a_z \sum_{\vec p \in \Omega} m_0 \bar \psi(\vec p) \psi(\vec p)$, the diagonal structure of $\gamma^0$ makes the eigenspinors even simpler; they are just 
\begin{equation}
u_0^1 = \left[ \begin{array}{c} 1 \\ 0 \\0\\ 0 \end{array}  \right], \quad u_0^2 = \left[ \begin{array}{c} 0 \\ 1 \\0\\ 0 \end{array}  \right], \quad v_0^1 = \left[ \begin{array}{c} 0 \\ 0 \\1 \\ 0 \end{array}  \right], \quad v_0^2 = \left[ \begin{array}{c} 0 \\ 0 \\0\\ 1 \end{array}  \right]. 
\end{equation}
The positive frequency eigenspinors for the physical Hamiltonian overlap with both of the negative frequency eigenspinors of the reference Hamiltonian. Thus
\begin{equation}
a^\alpha (p) = u^{\alpha \dagger}(p) \psi(\vec p) = u^{\alpha \dagger} u^\beta_0 a^\beta_0(\vec p) + u^{\alpha \dagger} v^\beta_0 b_0^{\beta \dagger}(\vec p), 
\end{equation}
\begin{equation}
b^{\alpha \dagger}(p) = v^{\alpha \dagger}(p) \psi(\vec p) = v^{\alpha \dagger} u^\beta_0 a^\beta_0(\vec p) + v^{\alpha \dagger} v^\beta_0 b_0^{\beta \dagger}(\vec p),
\end{equation}
where
\begin{equation}
u^{\alpha \dagger} u^\beta_0 = \sqrt{\frac{E+M}{2 E}} \delta^{\alpha \beta}, \quad v^{\alpha \dagger} v^\beta_0 = \sqrt{\frac{E+M}{2 E}} \delta^{\alpha \beta}, 
\end{equation}
\begin{equation}
u^{1 \dagger} v^1_0 = \frac{P_z}{\sqrt{2E(E+M)}}, \quad u^{1 \dagger} v^2_0 = \frac{P_x - i P_y}{\sqrt{2E(E+M)}},
\end{equation}
\begin{equation}
u^{2 \dagger} v^1_0 = \frac{P_x + i P_y}{\sqrt{2E(E+M)}}, \quad u^{1 \dagger} v^2_0 = - \frac{P_z}{\sqrt{2E(E+M)}},
\end{equation}
\begin{equation}
v^{1 \dagger} u^1_0 = -\frac{P_z}{\sqrt{2E(E+M)}}, \quad v^{1 \dagger} u^2_0 = \frac{-P_x + i P_y}{\sqrt{2E(E+M)}},
\end{equation}
\begin{equation}
v^{2 \dagger} u^1_0 = -\frac{P_x + i P_y}{\sqrt{2E(E+M)}}, \quad v^{1 \dagger} u^2_0 = \frac{P_z}{\sqrt{2E(E+M)}}.
\end{equation}
Using these relations, we find that the physical ground state is 
\begin{eqnarray} \label{4dt}
| 0000 \rangle_{phys} &=& \frac{1}{2E} \left[ (E+M) | 0000 \rangle_0 +  P_z | 0101 \rangle_0 - (P_x + i P_y) | 0110 \rangle_0  \right. \\ && \left. -(P_x-iP_y)  | 1001 \rangle_0-P_z  | 1010 \rangle_0 + (E-M) | 1111 \rangle_0 \right]. \nonumber
\end{eqnarray}
We want a unitary realising the transformation \eqref{4dt}. This is in a six-dimensional subspace of the Hilbert space $\mathcal H_p$, so we're looking for a $U(6)$ transformation $U_p$. The ambiguity in the choice of $U_p$ corresponds to left multiplication by a $U(5)$ transformation which fixes $|\vec 0 \rangle_{phys}$. Minimizing over this $U(5)$ ambiguity to find the $U_p$ closest to the origin in the standard metric on $U(6)$ is non-trivial; we have not carried out the calculation explicitly. 

However, we can simplify the problem considerably by noting that the momentum space Hamiltonian \eqref{4dH} is 
\begin{equation}
H = a_x a_y a_z \sum_{\vec p \in \Omega}\left[  M \bar \psi(\vec p) \psi(\vec p) + \sum_i P_i \bar \psi(\vec p) \gamma^i \psi(\vec p) \right] , 
\end{equation}
which looks just like the continuum Hamiltonian for a fermion of mass $M$ and momentum $\vec P$. As a result, we would expect the distance between the physical vacuum and the reference state in $\mathcal H_p$ to be invariant under the symmetries of a continuum theory, and depend only on $P^2$. If we make this assumption, we can determine the dependence on $P^2$ by considering a case with a single momentum component. 

For example, take just $P_z$. Then 
\begin{equation}
 u^1 = \frac{1}{\sqrt{2E(E+M)}} \left[ \begin{array}{c} M+E \\ 0 \\ P_z \\ 0 \end{array}  \right], \quad u^2 = \frac{1}{\sqrt{2E(E+M)}} \left[ \begin{array}{c} 0 \\ M+E \\ 0\\ - P_z \end{array}  \right], 
\end{equation}
with eigenvalue $E$, and
\begin{equation}
v^1 =  \frac{1}{\sqrt{2E(E+M)}} \left[ \begin{array}{c} -P_z \\ 0\\ M+E \\ 0  \end{array}  \right]\quad
 v^2 =  \frac{1}{\sqrt{2E(E+M)}} \left[ \begin{array}{c} 0\\ P_z \\0\\ M+E  \end{array}  \right],  \end{equation}
and $u^1, v^1$ are a mixture of $u^1_0, v^1_0$, while $u^2, v^2$ are a mixture of $u^2_0, v^2_0$. As a result, we can decompose the Hilbert space $\mathcal H_p$ as a tensor product of the space acted on by $a^1, b^1$ and the space acted on by $a^2, b^2$, and the physical ground state in $\mathcal H_p^1$ is 
\begin{equation} 
| 00 \rangle_{phys,1} =  \sqrt{\frac{E+M}{2E}}    | 00 \rangle_0  -\sqrt{\frac{E-M}{2E}}  | 11 \rangle_0 ,\nonumber
\end{equation}
while the physical ground state in  $\mathcal H_p^2$ is\footnote{To see that  $|0000 \rangle_{phys}  = | 00 \rangle_{phys,1} \times  | 00 \rangle_{phys,2}  = \frac{1}{2E} \left[ (E+M) | 0000 \rangle_0 +  P_z | 0101 \rangle_0 -P_z  | 1010 \rangle_0 + (E-M) | 1111 \rangle_0 \right]$, we need to recall that $|1111\rangle_0 = a_0^{1 \dagger} a_0^{2 \dagger} b_0^{1 \dagger} b_0^{2 \dagger} |0000\rangle_0$, so $|11 \rangle_0 \times |11 \rangle_0 = a_0^{1 \dagger} b_0^{1 \dagger} a_0^{2 \dagger} b_0^{2 \dagger} |0000 \rangle_0 = - |1111\rangle_0$. }
\begin{equation} 
| 00 \rangle_{phys,2} =  \sqrt{\frac{E+M}{2E}}    | 00 \rangle_0  + \sqrt{\frac{E-M}{2E}}  | 11 \rangle_0 .\nonumber
\end{equation}
Thus, the transformation from the reference state to the physical ground state is a product of a rotation in $\mathcal H_p^1$  and a rotation in $\mathcal H_p^2$. These individually are the same as the two-dimensional case. The complexity is then just the combination of the contributions from $\mathcal H_p^1$  and  $\mathcal H_p^2$. Thinking of this as calculated in the Riemannian metric on the unitaries on $\mathcal H_p$, the minimum distance would be obtained by rotating in the two factors simultaneously, giving $\mathcal C(U_p) = \sqrt{2} \theta$, where\footnote{If we combined  the contributions from $\mathcal H_p^1$  and  $\mathcal H_p^2$ in a Manhattan metric, we would have $\mathcal C(\vec p) = 2 \theta$. This seems less appropriate as the way $\mathcal H_p$ splits up into a tensor product depends on which component of the momentum we consider; for example taking $P_x$ decomposes $\mathcal H_p$ into a subspace acted on by $a^1, b^2$ and a subspace acted on by $a^2, b^1$. But this overall numerical difference is in any case unimportant for our considerations.}
\begin{equation} 
\cos \theta = \sqrt{\frac{ E+M}{2E}}. 
\end{equation}
We obtained this result by considering a momentum where only $P_z$ was non-zero, but if we assume the complexity is a function only of $P^2$, we can apply this result to all the momenta in the lattice. We can check that we get the same answer by taking a different component, that is taking only $P_x$ or $P_y$ nonzero. 

\begin{figure}
\centering 
\includegraphics[width=0.6\textwidth]{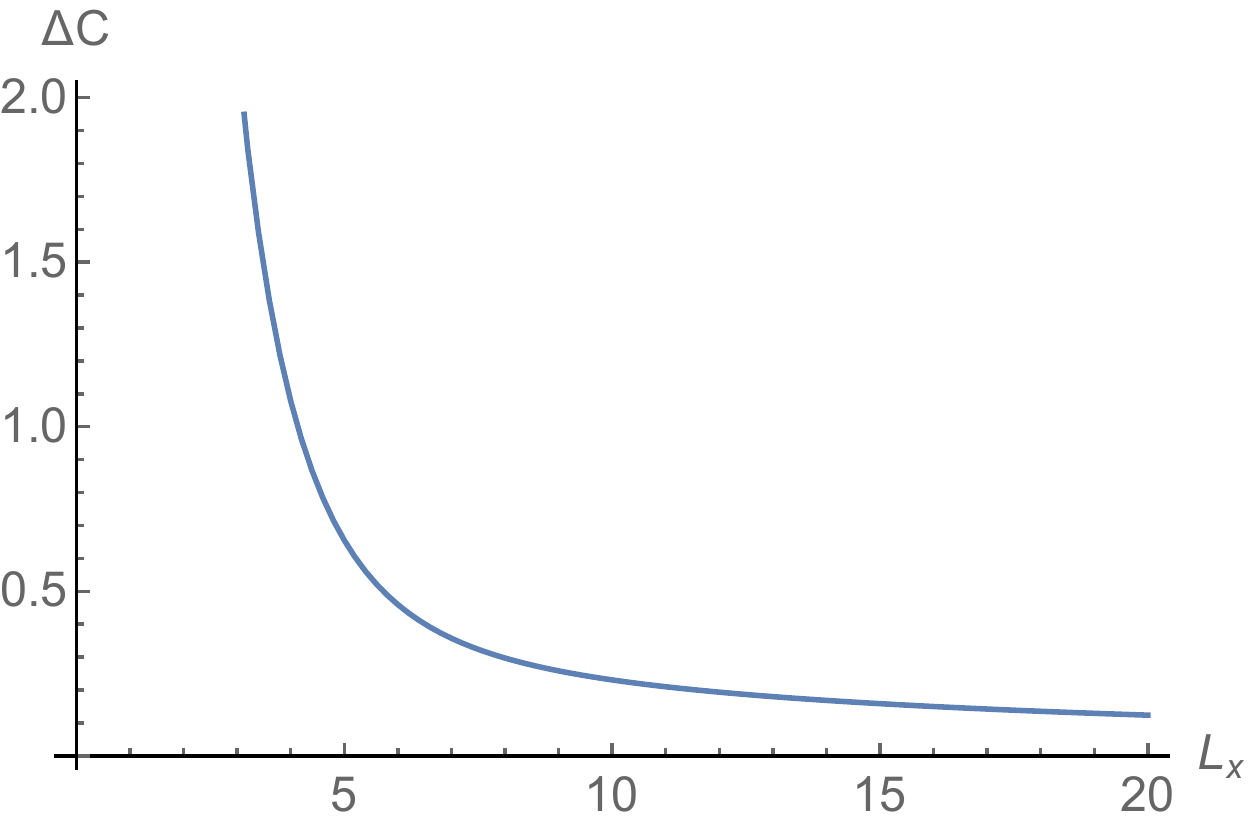}
\caption{We plot the difference $\Delta \mathcal C$ between the complexity of the ground states for a fermion with antiperiodic boundary conditions in one direction and a fermion with periodic boundary conditions in both directions on a three-dimensional spatial lattice, as a function of the size of the circle with the varying boundary conditions. We see that the difference is positive, and decreases as we increase the size of the circle.} \label{comp4d}
\end{figure}

We add up these contributions for each site in the momentum lattice, and take the difference between antiperiodic and periodic boundary conditions to calculate $\Delta \mathcal C$ in \eqref{deltac}. We plot this difference as a function of $L_x$ in figure \ref{comp4d}. We see that the difference is  again positive, and decreases as a function of the size of the circle. In this case this is also the behaviour expected holographically. Holographically, the complexity would fall off as $1/L_x^2$. Again the numerical results for the lattice computation look slightly faster than in three dimensions but are not well fit by a simple power law. 

We can calculate the complexity in higher dimensions along similar lines. As we increase the spacetime dimension the dimension of the spinor representation increases, so the calculation for generic momentum values gets more complicated, but we can proceed by doing the calculation in the case where the momentum has only one non-zero component, where the unitary transformation will again be a simple rotation, and extrapolating to the general case assuming the complexity at a given momentum is a function only of $P^2$. 

\bibliographystyle{JHEP}
\bibliography{complexity}

\end{document}